\documentclass[aps,prb,preprint,floatfix,superscriptaddress,showpacs,titlepage]{revtex4-2}
\usepackage{graphicx}% Include figure files
\usepackage{dcolumn}% Align table columns on a decimal point
\usepackage{amsmath,bm}% bold math
\usepackage{color}%
\usepackage{braket}
\usepackage{multirow}
\usepackage{array} % adjust table width 
\usepackage{xr-hyper}
\usepackage{url}
\usepackage{hyperref}
\usepackage[authormarkup=none]{changes}

\usepackage{kotex}

\definechangesauthor[color=purple]{BG}

\definechangesauthor[color=cyan]{SJ}

\newcommand*{\rom}[1]{\expandafter\@slowromancap\romannumeral #1@}
\makeatletter
\newcommand*{\addFileDependency}[1]{% argument=file name and extension
  \typeout{(#1)}
  \@addtofilelist{#1}
  \IfFileExists{#1}{}{\typeout{No file #1.}}
}
\makeatother
 
\newcommand*{\myexternaldocument}[1]{%
    \externaldocument{#1}%
    \addFileDependency{#1.tex}%
    \addFileDependency{#1.aux}%
}

\myexternaldocument{SI}

\begin{document}

\title{High Photovoltaic Efficiency in Bulk-Stacked One-Dimensional GeSe$_2$ van der Waals Crystal}

\author{Seoung-Hun Kang}
\thanks{These two authors contributed equally}
\affiliation{Department of Physics, Kyung Hee University, Seoul, South Korea}
\affiliation{Department of Information Display, Kyung Hee University, Seoul, South Korea}
\affiliation{Research Center for Technology Commercialization, Korea Institute of Science and Technology Information (KISTI), Seoul, South Korea}
\author{Youngjae Kim}
\thanks{These two authors contributed equally}
\affiliation{School of Physics, Korea Institute for Advanced Study (KIAS), Seoul 02455, Korea}
\affiliation{Department of Semiconductor Physics, Kangwon National University, Chuncheon 24341, Republic of Korea}
\author{Bo Gyu Jang}
\email[Corresponding author. Email:]{bgjang@khu.ac.kr}
\affiliation{Department of Materials Science and Engineering, Kyung Hee University, Yongin, Gyeonggi 17104, South Korea}
\author{Sejoong Kim}
\email[Corresponding author. Email:]{sejoong@hongik.ac.kr}
\affiliation{Department of Electronic and Electrical Convergence Engineering, Hongik University, Sejong 30016, South Korea}

\begin{abstract}

Germanium diselenide (GeSe$_2$) has recently attracted substantial interest as a rare example of one-dimensional (1D) van der Waals material. Here, we investigate the photovoltaic potential of bulk-stacked GeSe$_2$ chains using first-principles calculations within the $GW_{0}$ approximation and the Bethe–Salpeter equation (BSE) to capture quasiparticle and excitonic effects. The bulk GeSe$_2$ exhibits indirect $GW$ band gaps of 1.92~eV (type-I) and 1.08~eV (type-II). Optical calculations show markedly stronger visible-light absorption in type-II, yielding a spectroscopically limited maximum efficiency (SLME) of $\sim$25.6\% at a 0.5~$\mu$m thickness. Phonon and room-temperature ab initio molecular dynamics analyses indicate that type-II is dynamically stable, whereas type-I shows imaginary phonon modes, suggesting a propensity for structural distortion. These results identify type-II GeSe$_2$ as a promising stable absorber for thin-film photovoltaics with enhanced flexibility compared to typical 2D vdW systems.

\end{abstract}

\maketitle

%\textcolor{blue}{\textbf{To-do list
%\begin{enumerate}
%    \item Please specify band gap positions (ex. X or $\Gamma$)
%    \item Please summarize band gap energies in Table~2. 
%    \item Draw the irreducible Brillouin zone and high symmetry points
%    \item Do we use the same BZ boundaries for type-1 and type-2?
%    \item Please add the ab initio molecular dynamics results for thermal stability. 
%    \item Please summarize computational details. 
%\end{enumerate}
%}}

\section{Introduction}
In the past decade, van der Waals (vdW) low-dimensional materials have attracted widespread attention because their intrinsic electronic, optical, and topological properties differ markedly from those of conventional three-dimensional (3D) bulk compounds~\cite{Sci2011,ACS2013,NN2012,CR2017,NN2010,PRL2010,Nat2013,Sci2004}. 
Unlike 3D solids, where strong covalent or ionic bonds dominate, vdW systems consist of discrete building blocks such as two-dimensional (2D) layers or one-dimensional (1D) chains held together by relatively weak inter-component forces known as vdW interactions~\cite{Wilson1975,Novoselov2005,Chhowalla2013,Manzeli2017,Friend1987,Wilson1969,Wang2018,Heine2015}. 
This weak vdW binding makes it possible to exfoliate individual sheets or chains~\cite{Sci2004} and to stack different layers with atomic precision~\cite{Dean2010, CastellanosGomez2014}. 
These capabilities enable designer heterostructures whose optoelectronic behavior can be tuned on demand~\cite{Nat2013,NN2012}.

Much of the early work focused on 2D vdW materials, especially graphene~\cite{Geim2007} and transition metal dichalcogenide monolayers~\cite{ACS2013}. 
%
%Even for 1D vdW cases, 
Recent studies have begun to explore truly 1D vdW crystals that naturally form as isolated chains rather than being carved from a parent-layered bulk~\cite{Xiang2020}.
Classic examples such as carbon nanotubes predate graphene and already showcase exceptional mechanical, electrical, and thermal performance~\cite{Ijima1991}. 
% SJK: 나노튜브는 그래핀에서 유도되어 만들어지는 것으로 봐야 하지 않을까 싶습니다. 시기적으로는 그래핀보다 먼저 나왔지만, 합성 방법이나 발견 방법의 시기적인 선후관계에 의한 것이지 않을까요? 그냥 둬도 무방합니다. 
%
Quasi-1D nanoribbons derived from 2D sheets have also been investigated for nanoelectronics and quantum devices~\cite{Son2006}. 
Intrinsically 1D vdW materials, such as selenium or tellurium chains, also exist, without being synthesized from layered precursors~\cite{Sleight1972}. 
%
% SJK: Ref. 24 논문이 존재하지 않습니다. 다른 논문으로 대체하였습니다. 
These materials present new opportunities for fundamental research and technological applications, offering distinct electronic, optical, and topological characteristics not accessible in 2D or 3D frameworks~\cite{BALANDIN202274}.
%
%Intrinsically 1D vdW systems, for instance, selenium or tellurium chains~\textbf{[REFERENCES HERE]}, open a distinct avenue. 
%
%In 1D, reduced dielectric screening amplifies electron–electron interactions and can produce novel optical and topological phenomena not accessible in 2D or 3D frameworks~\cite{Zhang2020}.

%   

%\addbg{여기서는 GeSe2에만 너무 집중하지 않고, Si or Ge dichalcogenide family에 대해 더 브로드하게 작성해도 괜찮지 않을까요? 괜찮으시면 이 문단의 GeSe2를 제너럴하게 바꾸고 다음 문단에서 GeSe2 chain gap이 soloar spectrum window에 맞으니 GeSe2에 집중한다로 넘어가도 될 것 같습니다.}
%\addsj{SJK: 좋습니다. 그렇게 바꿔볼까요?}
% Silicon or germanium dichalcogenides (AX$_2$, A = Si or Ge and X = S or Se)
{Germanium diselenide (GeSe$_2$)} exemplifies this class of true 1D vdW materials.
Each Ge atom coordinates four Se atoms to form a tetrahedral GeSe$_4$ unit. 
The adjacent tetrahedra share a two-Se edge to produce an extended chain-like lattice~\cite{Lee2023, doi:10.1021/acsnano.4c04184}. 
Recently, an alternative GeSe$_2$ chain topology was synthesized inside carbon nanotube templates, demonstrating that GeSe$_2$ can adopt multiple 1D conformations at the nanoscale~\cite{Lee2023, doi:10.1021/acsnano.4c04184}. 
%
% SJK: type-I과 type-II에 대한 정의를 여기에서 언급합니다. 
Both the naturally occurring and template-assisted GeSe$_2$ chains, which are called type-I and type-II, respectively, exhibit semiconducting behavior. 
%
% type-II에 대한 PBE bandgap만 있어서 type-I와 type-II를 다 언급하였습니다. 
Standard density functional theory (DFT) predicts that the type-I and type-II isolated GeSe$_2$ chains possess mean-field bandgaps of about 1.90 and 0.80~eV, respecitvely~\cite{Lee2023, doi:10.1021/acsnano.4c04184}.

However, in 1D systems, reduced screening amplifies many-body effects that are not captured by mean-field DFT. 
Quasiparticle gaps typically increase when computed using the $GW$ approximation~\cite{Hedin1965}, and strong excitonic binding can further reshape optical spectra~\cite{Spataru2004}. 
%
% BGJ: 위에서 언급한대로 윗문단은 제너럴하게 AX2 family에 대해서 논하는게 괜찮다면
% Among AX$_2$ family, 
Since the GeSe$_2$ chain gap lies within the solar-spectrum window, these materials could be promising for next-generation photovoltaics~\cite{Tsakalakos2007}. 
A reliable assessment of their optoelectronic performance requires an accurate treatment of electron–electron interactions.

In this study, we provide a comprehensive computational investigation to demonstrate that the stacked bulk structure of GeSe$_{2}$ chains holds significant potential for photovoltaic applications. 
Given the reduced dimensionality of the constituent 1D GeSe$_{2}$ chains, an accurate consideration of electron-electron interactions is essential to obtain reliable electronic and optical properties. 
To address this issue, we perform ab initio many-body perturbation theory calculations within the $GW_{0}$ approximation, which go beyond the limitations of conventional mean-field DFT calculations. 
This approach provides a more accurate description of quasiparticle excitations and electronic screening effects, both of which play a crucial role in determining the optoelectronic behavior of the material. 
By analyzing the optical spectra of the GeSe$_2$ chains, we further assess their photovoltaic efficiency using the spectroscopically limited maximum efficiency (SLME) metric~\cite{PhysRevLett.108.068701}. 
Our SLME calculations reveal that the GeSe$_{2}$ chain structure can achieve a photovoltaic efficiency of up to 25.63\% when stacked to a thickness of 0.5 $\mu$m. 
This efficiency is comparable to or even exceeds that of well-established vdW 1D photovoltaic materials, Sb$_2$S$_3$ and Sb$_2$Se$_3$,~\cite{KONDROTAS2018857,MAVLONOV2020227}, highlighting the potential of GeSe$_{2}$-based nanostructures for future solar energy conversion.

\section{Computational Details}
All first‐principles calculations were performed using the Vienna \textit{Ab initio} Simulation Package (VASP)~\cite{Kresse1996,Kresse1999} with the projector-augmented wave method (PAW)~\cite{Blochl1994}. 
The exchange–correlation potential was treated within the generalized gradient approximation (GGA) in the Perdew–Burke–Ernzerhof (PBE) form~\cite{PBE1996}. 
We used the DFT-D3 dispersion correction to describe the inter-chain van der Waals interactions~\cite{GrimmeD3}, and also verified the key results using the revised vdW-DF2 functional with B86R exchange (rev-vdW-DF2, also referred to as vdW-DF2-B86R)~\cite{PhysRevB.89.121103, PhysRevB.95.180101}, where the revised Becke exchange functional (B86R)~\cite{10.1063/1.451353} is adopted for the exchange functional and a non-local correction is described by the second version of nonlocal vdW-DF (vdW-DF2)~\cite{PhysRevLett.92.246401, PhysRevB.82.081101}.  
A plane‐wave energy cutoff of 500~eV was used for expanding the Kohn–Sham wavefunctions. 
The Brillouin zone (BZ) integrations used the $\Gamma$-centered $15\times15\times15$ Monkhorst-Pack $k$-grid for structural relaxations and total energy calculations. 
All atomic positions and lattice parameters were relaxed until the residual Hellmann–Feynman forces on each atom were below 0.01~eV/$\AA$ and the total energy change between successive ionic steps was less than $10^{-5}$~eV.

Starting from the ground state converged DFT-GGA, the quasiparticle energies were obtained using the single-shot $G_0W_0$ approximation with 
%partial self‐consistency 
self-consistent update for $G$, which is denoted by $GW_0$~\cite{Hybertsen1986,Shishkin2006}.
%in $G$ (denoted $GW_0$)~\cite{Hybertsen1986,Shishkin2006}. 
In the $GW_0$ calculations, 180 bands (occupied and unoccupied) were included to ensure the convergence of the self‐energy. The screened Coulomb interaction $W$ was treated within the plasmon‐pole approximation~\cite{Hybertsen1986}, and the dielectric matrix was constructed on an $8\times8\times8$ $k$‐point grid with a plane‐wave cutoff of 211~eV for the response‐function expansion.

To incorporate excitonic effects in the optical spectra, we solved the Bethe–Salpeter equation (BSE) on top of the $GW_0$ quasiparticle energies~\cite{Rohlfing1998a,Albrecht1998,Rohlfing2000,Onida2002,Benedict1999}. 
The BSE Hamiltonian included the top 12 valence bands and the lowest 8 conduction bands. 
The electron–hole kernel was constructed using the same $8\times8\times8$ $k$‐point sampling, retaining local‐field effects. 
From the BSE solution, we extracted the imaginary part of the macroscopic dielectric function $\varepsilon_{2}$$(\omega)$ for photon energies up to 10.0~eV. 
The resulting $GW_0$+BSE absorption spectrum was compared to the independent‐particle result at the $GW_0$ level---denoted by $GW$+RPA---to assess the impact of electron–hole interactions on optical properties~\cite{Benedict1999}.

\section{Results and Discussion}

%===========================================================
\begin{figure}[t]
\includegraphics[width=1.0\columnwidth]{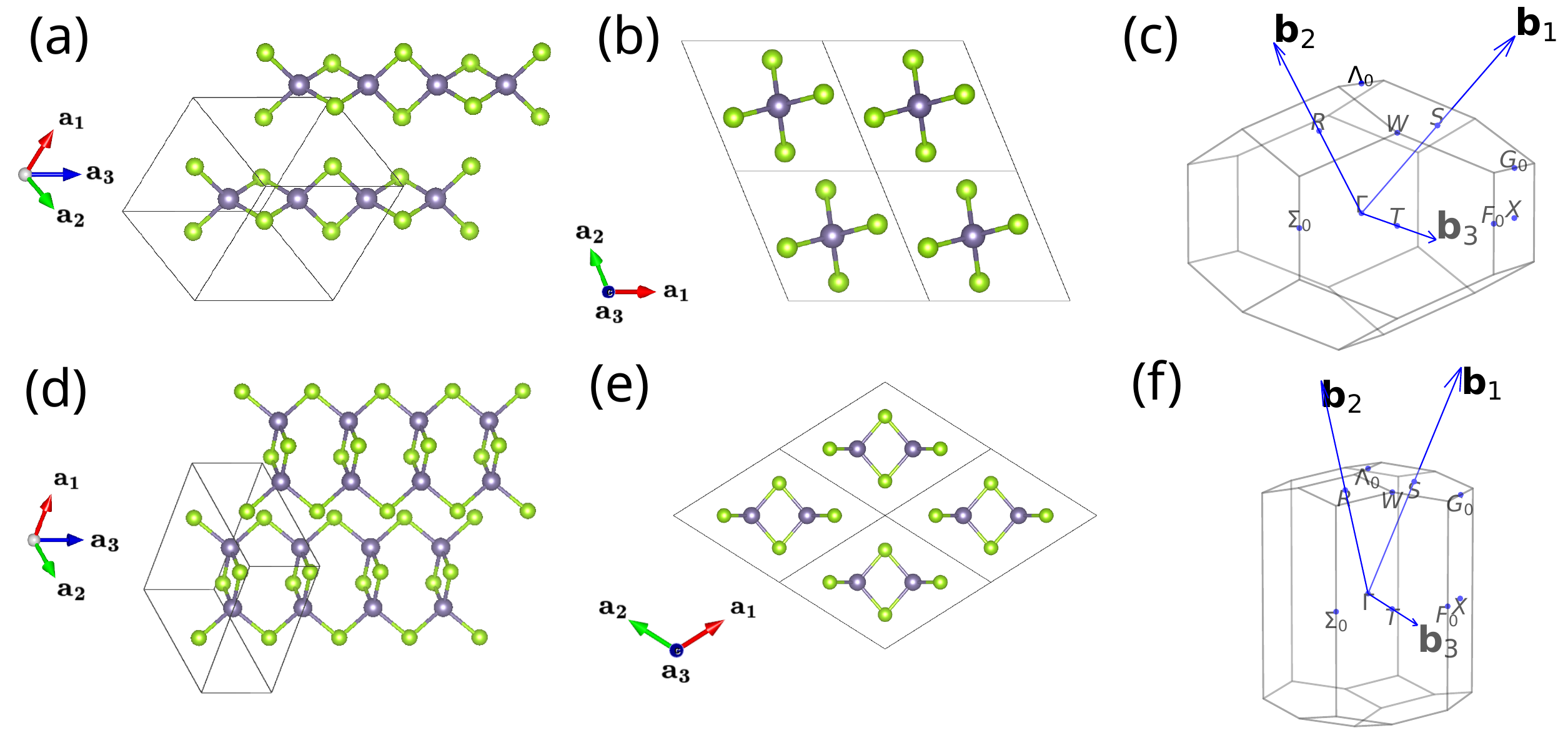}
\caption{\label{FIG1} (Color online) Atomic configurations of (a,b) type-I bulk GeSe$_{2}$ and (d,e) type-II bulk GeSe$_{2}$. Triclinic primitive unit cells and corresponding lattice vectors $\mathbf{a}_{1}$ (red), $\mathbf{a}_{2}$ (green) and $\mathbf{a}_{3}$ (blue) are illustrated with atomic chains. Corresponding irreducible Brillouin zones (IBZ) are illustrated for (c) type-I and (f) type-II chains. Reciprocal lattice vectors $\{\mathbf{b}_{1}, \mathbf{b}_{2}, \mathbf{b}_{3}\}$ and high-symmetric points are displayed together. Atomic configurations are drawn by using VESTA~\cite{Momma:db5098}.}
\end{figure}
%=============================================================
%
Figure~\ref{FIG1} shows the atomic structures of two bulk polymorphs of GeSe$_{2}$, each formed by stacking one-dimensional chains via van der Waals forces. 
In type‐I GeSe$_{2}$ (Fig.~\ref{FIG1}(a)), each germanium atom is coordinated to four selenium atoms, forming a GeSe$_{4}$ tetrahedron. 
Neighboring tetrahedra share a two‐selenium edge, resulting in a continuous 1D tetrahedral chain. 
In contrast, the recently synthesized type‐II GeSe$_{2}$ chain is built from pairs of edge‐sharing tetrahedra, as shown in Fig.~\ref{FIG1}(b). 
These pairs repeat along the chain direction by sharing only selenium corners, producing a different connectivity pattern.

When these 1D chains are assembled into a bulk phase via vdW interactions, both type‐I and type‐II structures adopt triclinic unit cells. 
The lattice constants $a_{i}$ ($i=1,2,3$) and the interaxial angles $\theta_{ij}$ between the vectors $\mathbf{a}_{i}$ and $\mathbf{a}_{j}$ are listed in Table~\ref{Table1}. 
In each case, the chains run parallel to $\mathbf{a}_{3}$, and the in‐plane lattice constants satisfy $a_{1} = a_{2}$. 
Together, these parameters fully describe the triclinic geometry of the bulk GeSe$_{2}$ phases derived from their respective 1D chains.

%
%Figure~\ref{FIG1} illustrates the atomic structures of bulk type-1 and type-2 GeSe$_{2}$, where 1D chains are stacked via the vdW force. 
%
%The type-1 GeSe$_2$ chain is a 1D tetrahedral chain, where a germanium atom and four surrounding chalcogens Se form a tetrahedron, and two neighboring tetrahedra share a two-chalcogen edge, as shown in Fig.~\ref{FIG1}(a). 
%
%The type-2 GeSe$_2$ chain, synthesized in the recent experiment, is composed of a building block of two tetrahedra sharing a chalcogen edge, and building blocks are repeated in the chain direction with chalcogen corner-sharing, as shown in Fig.~\ref{FIG1}(b).  
%
%When the two types of GeSe$_{2}$ chains are stacked in a bulk phase, the primitive unit cells for both cases are triclinic. 
%
%Lattice parameters are summarized in Table~\ref{Table1}, where $a_{i}$ ($i=1,2,3$) is the lattice constant and $\theta_{ij}$ is the angle between lattice vectors $\mathbf{a}_{i}$ and $\mathbf{a}_{j}$. 
%
%Note that the 1D GeSe$_{2}$ chain is aligned along $\mathbf{a}_{3}$, and $a_{1}=a_{2}$.

%===========================================================
\begin{table}[b]
\small
  \caption{Lattice parameters for type-I and type-II bulk GeSe$_{2}$.}
  \label{Table1}
  \begin{tabular*}{0.48\textwidth}{@{\extracolsep{\fill}}lll}
    \hline
    \hline
    lattice parameters & type-I & type-II \\
    \hline
    \hline
    $a_{1}$          & 6.3494 \AA &  7.2645 \AA \\
    $a_{2}$          & 6.3494 \AA &  7.2645 \AA \\
    $a_{3}$          & 6.3494 \AA &  7.2645 \AA \\
    $\theta_{12}$    & 121.825$^\circ$ &  117.949$^\circ$\\
    $\theta_{23}$    & 86.889$^\circ$ &  70.191$^\circ$\\
    $\theta_{31}$    & 121.799$^\circ$ &  150.482$^\circ$\\
    \hline
  \end{tabular*}
\end{table}
%===========================================================

%===========================================================
\begin{figure}[t]
\includegraphics[width=1.0\columnwidth]{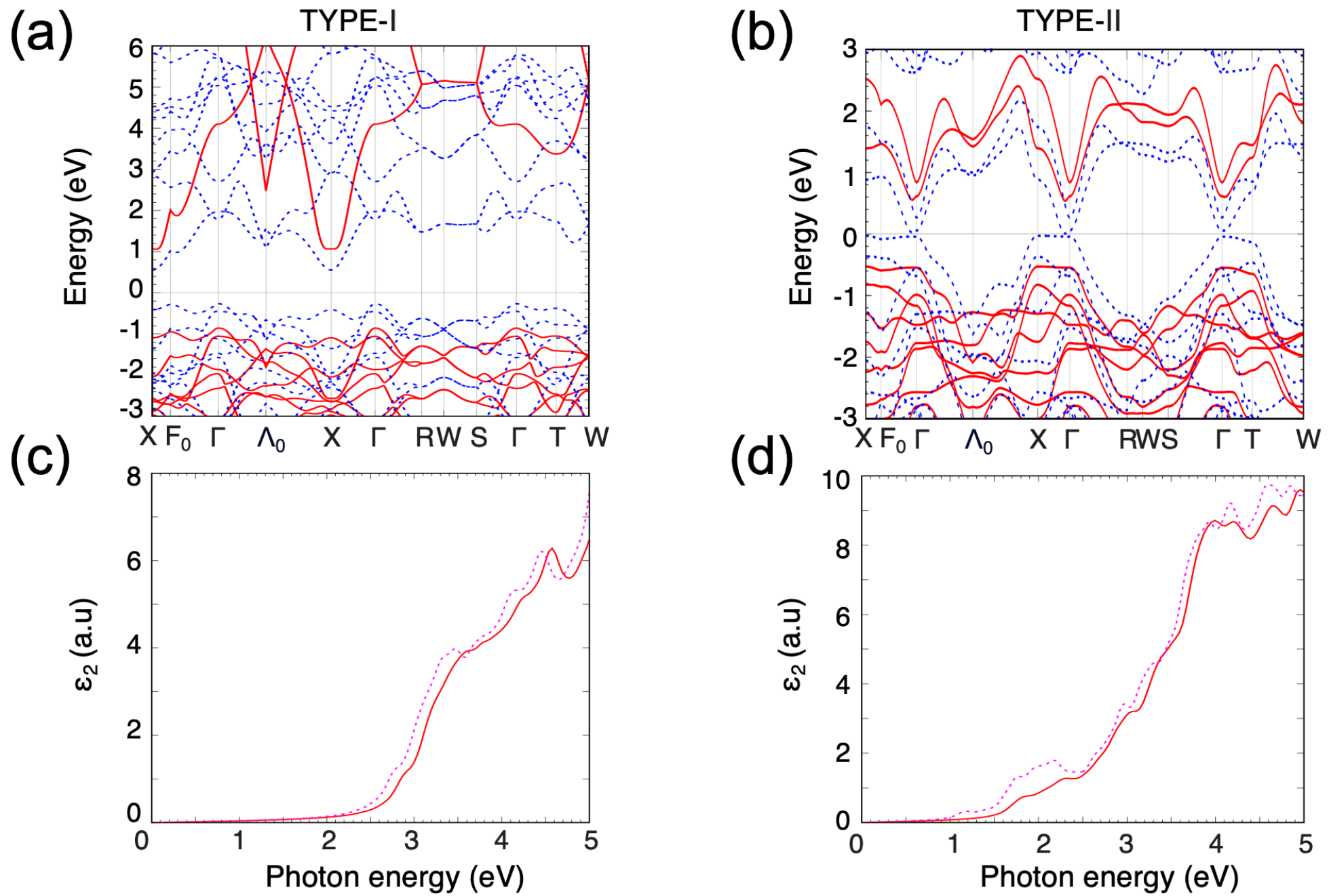}
\caption{
\label{FIG3} Electronic structure of bulk type-I and type-II GeSe$_{2}$. Electronic band structures computed by DFT-GGA (blue dashed) and $GW$ (red solid) approximations for (a) type-I bulk GeSe$_{2}$ and (b) type-II bulk GeSe$_{2}$. Absorption spectra $\varepsilon_{2}$ from $GW$ + BSE (magenta dashed) and $GW$ + RPA (red solid ) for (c) type-I bulk GeSe$_{2}$ and (d) type-II bulk GeSe$_{2}$.
%\addbg{Different color scheme for (c) and (d)? figure (a) (b)에서 파란색 점선이 DFT라서 (c)와 (d) 매칭이 혼란을 줄 수 도 있을 것 같습니다. 어떠신가요?}\addsj{네, 파란색 dotted line을 파란색 dashed line으로 바꾸는게 좋겠습니다.}\addsj{type-I의 GW의 conduction band가 다 나오도록 그려주실 수 있으신가요?} 
}
\end{figure}
%=============================================================
% figure (a) (b)에서 파란색 점선이 DFT라서 (c)와 (d) 매칭이 혼란을 줄 수 도 있을 것 같습니다. 어떠신가요?

Accurate band‐gap values are essential for assessing the photovoltaic potential of GeSe$_{2}$. 
To this end, we computed the band structures of both type‐I and type‐II bulk GeSe$_{2}$ using three approaches: DFT‐GGA, the HSE06 hybrid functional~\cite{JChemPhys_118_8207_2003, JChemPhys_125_224106_2006}, and the $GW_{0}$ approximation. 
Figures~\ref{FIG3}(a) and (b) represent the DFT‐GGA and $GW_{0}$ band structures for type‐I and type‐II GeSe$_{2}$, respectively. 
Both polymorphs have an indirect fundamental gap located between the \textbf{$\Gamma$} and \textbf{$X$} points. 
%In both polymorphs, the fundamental gap is indirect, which is located at \textbf{$\Gamma$} and \textbf{$X$}, respectively. 
%
The calculated band gaps are summarized in Table~\ref{Table2}.

%
%When compared to the previously reported DFT‐GGA band gaps of isolated 1D GeSe$_{2}$ chains (1.91~eV for type‐1 and 0.79~eV for type‐2~\cite{Lee2023}), the bulk‐stacked structures exhibit significantly smaller gaps at the GGA level. 
%
When compared to the DFT‐GGA band gaps of isolated 1D GeSe$_{2}$ chains, 1.91~eV for type‐I and 0.79~eV for type‐II~\cite{Lee2023}, the bulk‐stacked structures exhibit significantly smaller gaps at the GGA level as shown in Table~\ref{Table2}. 
In particular, type‐II bulk GeSe$_{2}$ shows an extremely small GGA gap of 0.09~eV, indicating that 3D stacking enhances dielectric screening and consequently reduces the mean‐field gap in comparison to low-dimensional cases. 

%
%Hybrid functional calculations using HSE06 increase these GGA gaps by roughly 0.9~eV in both cases. 
%
%Nonetheless, even HSE06 underestimates the true quasiparticle gap. 
%
%\textbf{[SJK:지금 보니까 HSE06 밴드갭이 type 2에서는 GW와 거의 비슷하네요? 0.08 eV 차이인데 underestimate라고 해도 괜찮을까요? 수치적으로 작은 것은 사실인데 그 차이가 크지 않아서.]}
%
%In the $GW_{0}$ approximation—where electron–electron interactions and screening are treated explicitly—the %band gaps open to 1.92~eV for type‐I and 1.08~eV for type‐II. 

%
Advanced calculations beyond the mean-field level yield band gaps larger than the GGA values.
Compared to DFT-GGA calculations, HSE06 calculations enhance band gaps by roughly 0.9 eV in both bulk polymorphs. 
The $GW_{0}$ approximation predicts even larger band gaps.
While for type-I GeSe$_{2}$, the $GW_{0}$ approximation increases the band gap to 1.92 eV, it produces a gap for type-II GeSe$_{2}$ (1.08 eV) slightly larger than the HSE06 value (1.00 eV).
%
%\delsj{For type-II GeSe$_{2}$, the HSE06 gap (1.00 eV) is already close to the $GW_{0}$ value (1.08 eV), while for type-I GeSe$_{2}$ $GW_{0}$ further increases the gap to 1.92 eV}. 
%
%\delsj{Therefore, we use $GW_{0}$ quasiparticle energies consistently for subsequent optical (BSE) and SLME analyses}.
%
Figure~\ref{FIG3} also shows that many‐body corrections push the valence bands downward and the conduction bands upward compared to the mean‐field result.  

We use $GW_{0}$ quasiparticle energies consistently for subsequent optical analyses.
The optical absorption spectra were obtained by solving the Bethe–Salpeter equation (BSE) based on $GW_{0}$ quasiparticle energies~\cite{Rohlfing1998a, Albrecht1998, Rohlfing2000, Onida2002, Benedict1999}.  
%To obtain the optical absorption spectra, we solved the Bethe–Salpeter equation (BSE) using the $GW_{0}$ quasiparticle energies~\cite{Rohlfing1998a, Albrecht1998, Rohlfing2000, Onida2002, Benedict1999}. 
%An $8\times8\times8$ $k$‐point grid and 12 valence and 8 conduction bands were used to ensure convergence of the BSE Hamiltonian. 
%
From the BSE solution, we extracted the imaginary part of the dielectric function, $\varepsilon_{2}(\omega)$, up to photon energies of 10.0~eV. Figures~\ref{FIG3}(c) and (d) compare the absorption spectra obtained with ($GW$+BSE) and without ($GW$+RPA) electron–hole interactions for type‐I and type‐II GeSe$_{2}$. 
%
%{\color{cyan}[SJK:type-I에서 어느 부분을 pronounced excitonic peak이라고 보는게 좋을까요?]}
The inclusion of electron–hole coupling through the BSE reveals excitonic features that are absent in $GW$+RPA spectra. For type-I, the $GW$+BSE calculation does not show low-energy excitonic peaks, whereas type-II exhibits a distinct excitonic peak around 2 eV. These results demonstrate that many‐body effects are crucial for accurately modeling both the electronic band gaps and the optical response of bulk GeSe$_2$ in the context of photovoltaic applications.

\begin{table}[b]
\small
  \caption{\ Indirect (fundamental) band gap energies $E_{g}$ for type-I and type-II bulk GeSe$_{2}$ calculated by DFT+GGA, hybrid functional HSE06, and $GW_{0}$ approximations.}
  \label{Table2}
  \begin{tabular*}{0.48\textwidth}{@{\extracolsep{\fill}}lll}
    \hline\hline
    calculation method & type-I & type-II \\
    \hline\hline
    DFT+GGA      & 0.80 eV & 0.09 eV \\
    HSE06        & 1.69 eV & 1.00 eV \\
    $GW_{0}$         & 1.92 eV & 1.08 eV \\
    % $GW_{0}$     & 1.85 eV & 1.25 eV \\
    \hline\hline
  \end{tabular*}
\end{table}
%=============================================================

%===========================================================
\begin{figure}[t]
\includegraphics[width=1.0\columnwidth]{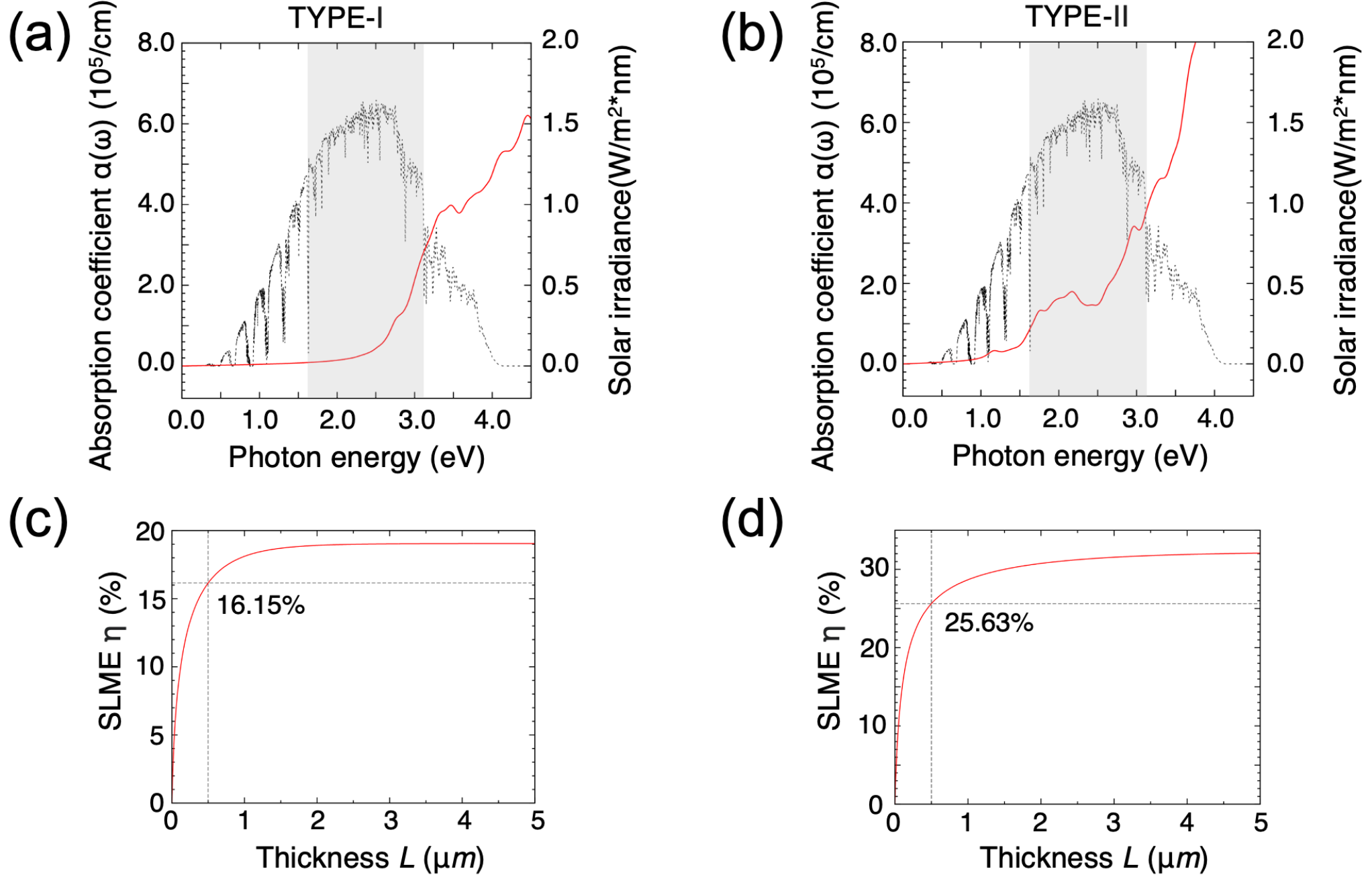}
\caption{\label{FIG4}Absorption coefficients (red solid line) of (a) type-I and (b) type-II bulk GeSe$_{2}$. The solar irradiance spectra, Air Mass 1.5, is shown with gray lines. The gray box specifies the energy range of visible light. The SLME $\eta$ of (c) type-1 and (d) type-2 bulk GeSe$_{2}$ are drawn as a function of thickness. For thickness 0.5 $\mu$m, the values of SLME are explicitly indicated.}
\end{figure}
%=============================================================

%
Figures~\ref{FIG4}(a) and (b) depict the calculated absorption coefficients $\alpha(\omega)$ of type-I and type-II GeSe$_2$ structures, respectively, overlayed with the standard AM 1.5 solar irradiance spectra~\cite{AM1.5}. 
Remarkably, type-II GeSe$_2$ shows a substantially stronger absorption within the visible spectral range (highlighted by gray shading in Fig.~\ref{FIG4}) than its type-I counterpart. 
This enhanced absorption indicates a superior capability for photovoltaic energy conversion, attributed to optimal electronic transitions directly aligned with the energies of visible photons.

To quantitatively assess the photovoltaic potential, we computed the SLME metric~\cite{PhysRevLett.108.068701}, which provides realistic efficiency estimates by considering intrinsic optical absorption properties, electron-hole interactions, and recombination mechanisms. 
Figures~\ref{FIG4}(c) and (d) illustrate the SLME $\eta$ as a function of absorber thickness $L$ for both GeSe$_2$ types. 
%
% 0.5 micrometer가 practical thickness인가요? 0.5 micrometer가 표준이 된 이유가 있나요?
At a practical device thickness of 0.5 $\mu$m, type-I GeSe$_2$ demonstrates an SLME of approximately 16.15\%, whereas type-II achieves an even higher efficiency of about 25.63\%. 
The notably higher SLME for type-II GeSe$_2$ underlines its exceptional promise as a photovoltaic absorber.

%
%\textbf{[SJK:이 문장은 저와 강승훈 박사님 논문을 인용하는 것인데 이 물질 말고도 더 좋은 photovoltaic material이 많으니 굳이 넣지 않는게 어떻겠습니까? 같은 vdW 1D 물질의 photovoltaic efficiency와 비교하는게 좋겠습니다. 아니면 저희 것만 인용하는 것보다는 더 다양한 물질의 efficiency와 비교하면서 저희 논문을 살짝 넣는게 어떨까요?]}
%\textcolor{blue}{
%The efficiency of type-II GeSe$_2$ compares favorably to recently reported photovoltaic absorbers, including layered Si$_3$O (~27\% at 0.5 $\mu$m thickness)~\cite{Kim2020Si3O} and delafossite-Cu$_2$ZnSnO$_4$ (~28.2\% at 2.0 $\mu$m thickness)~\cite{CZTO2023}, both regarded as state-of-the-art materials with substantial efficiency potential.}
%
%Moreover, compared to previously studied 1D chain-based bulk photovoltaic materials, such as Sb$_2$Se$_3$ exhibiting an SLME of approximately 28.0\% at 2.0 $\mu$m thickness~\cite{Yu2015}, and Sb$_2$S$_3$ reaching an SLME of around 25\% at similar thickness~\cite{Li2019}, type-II GeSe$_2$ clearly emerges as a competitive candidate within this promising class of photovoltaic absorbers.

The predicted SLME of type-II GeSe$_2$ (25.63\% at 0.5~$\mu$m thickness) places it in the upper range of efficiencies reported in recent SLME-based assessments of lead-free absorbers~\cite{PhysRevLett.108.068701}. Similarly, computational screenings have suggested that some other lead-free candidates, such as layered Si$_3$O and delafossite-type Cu$_2$ZnSnO$_4$, may reach comparably high SLME values~\cite{Kim2020Si3O,CZTO2023}.
% {\color{red}[SHK:  SLME를 처음 언급하는 부분에 넣기에도 이상한 느낌입니다. 지금처럼 고쳤습니다.]}

%{\color{cyan}[SJK: 저희 논문을 인용하는 부분인데 문맥상 필요없는 부분이어서 어색해 보입니다. 그냥 차라리 SLME를 처음 언급하는 부분에서 인용하는게 어떨까요?]}
%\delsj{For context, several layered or complex oxides have also been predicted to reach similarly high SLME values in computational screenings (e.g., layered Si$_3$O and delafossite Cu$_2$ZnSnO$_4$)}~\cite{Kim2020Si3O,CZTO2023}. 
Since GeSe$_2$ is a bulk-stacked quasi-1D van der Waals material, a more direct comparison can be made with quasi-1D ribbon/chain chalcogenides such as Sb$_2$Se$_3$ and Sb$_2$S$_3$, which have been widely investigated as thin-film photovoltaic absorbers~\cite{MAVLONOV2020227,KONDROTAS2018857}. 
In particular, Sb$_2$S$_3$ and Sb$_2$Se$_3$ are reported to exhibit SLME values of approximately 23.5\% and 29.0\%, respectively, for a 0.5~$\mu$m film thickness~\cite{IEEE}, indicating that type-II GeSe$_2$ is competitive within this broader class of quasi-1D van der Waals photovoltaic absorbers.
%{\color{cyan}[SJK:인용된 Sb2Se3논문을 보니 두께에 따른 SLME을 계산해놓은 그림 Fig.~1(c)가 있습니다. 여기에서 0.5 micrometer에서의 Sb2Se3와 Sb2S3의 SLME을 읽으면 될 것 같습니다. ]}

% SJK: 이 문단은 앞의 내용을 요약하는 것으로 results보다는 conclusion 혹은 summary쪽에 가는 것이 좋겠습니다. 
%Overall, our findings demonstrate the considerable potential of GeSe$_2$ as a highly efficient photovoltaic absorber, particularly highlighting the advantageous optical and electronic characteristics of its type-II phase. This study provides a theoretical foundation for further experimental exploration and optimization of GeSe$_2$-based solar cells, emphasizing the continued importance of computational screening in the discovery of next-generation photovoltaic materials.

%===========================================================
\begin{figure}[t]
\includegraphics[width=1.0\columnwidth]{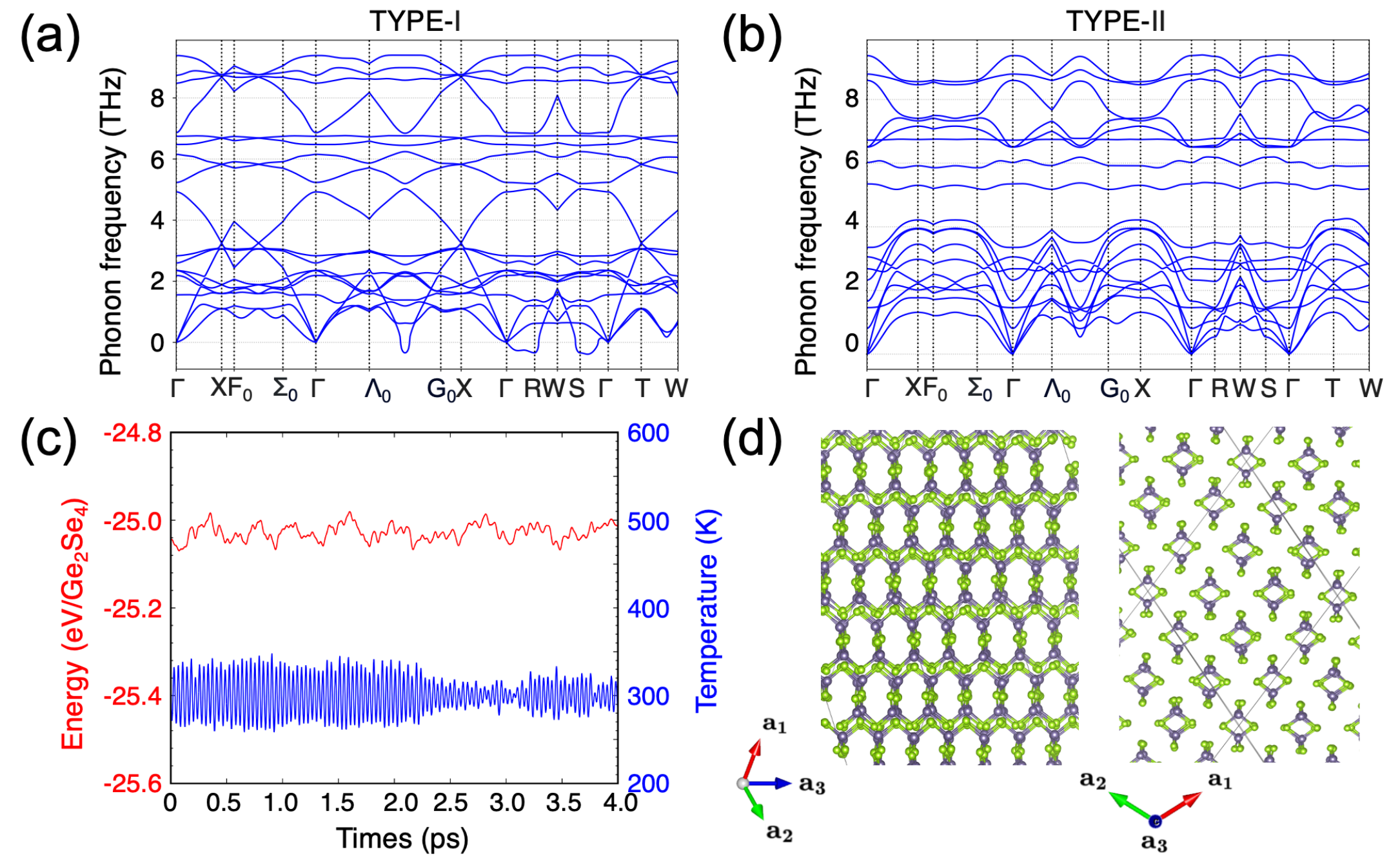}
\caption{\label{FIG5}Phonon frequencies of (a) type-I and (b) type-II bulk GeSe$_{2}$. (c) Total potential energy and temperature as a function of time during canonical MD simulations at 300 K. (d) The final structure at the end of the simulation time of 4 ps (inset).}
\end{figure}
%=============================================================
%
Having established the excellent photovoltaic efficiency and favorable optoelectronic characteristics of type-II GeSe$_2$, it is crucial to verify its structural and thermal stability, as practical applications require materials to retain their properties under realistic operating conditions. 
To this end, we first assessed dynamic stability by calculating the phonon dispersion spectra for type-I and type-II GeSe$_2$, as shown in Figs.~\ref{FIG5}(a) and (b). 
The phonon dispersion of type-I GeSe$_2$ exhibits imaginary (negative) frequencies in parts of the Brillouin zone, suggesting dynamical instability and a tendency toward structural distortions. By contrast, type-II GeSe$_2$ shows no imaginary frequencies across the Brillouin zone, supporting its dynamical stability. 
% SJK: phonon 성질을 연구하는 논문은 아니라서 이 부분은 comment out처리하겠습니다. 
%Additionally, we observed anisotropic slopes of the longitudinal acoustic (LA) phonon branches near the $\Gamma$ point, indicating direction-dependent mechanical rigidity along the $\Gamma–F_0$ and $\Gamma–\Lambda_0$ paths.
%{\color{red}{FIG.4 type-I 그림에서 negative frequency 가 figure상으로는 보기가 어려운데ㅣ넵 알겠습니다.}}
%\addsj{SJK:아, 제가 포논 결과를 업데이트 안해서 그래요. 포논 결과 다시 업데이트할게요.}

%
%While phonon dispersion analysis provides important information on harmonic structural stability at 0 K, it does not account for potential structural changes induced by thermal fluctuations at finite temperatures. 
%{\color{cyan}[SJK:total energy과 temperature time evolution을 둘다 그리는게 어떨까요? 현재 total energy의 y축이 간격이 작아서 fluctuation이 커보입니다. y축의 범위를 늘려서 fluctuation이 작게 보이도록 해주시고, 한 그래프 안에 temperature evolution도 같이 그려주세요. inset에 들어가 있는 snapshot은 작아서 잘 보이지 않네요. 혹시 fluctuation 그래프는 (c)로 그려주시고, (d)에 snapshot을 따로 그리는 것은 어떨까요? (a), (b), (c), (d)를 2x2 구조로 그리는 것입니다.]}
Although phonon calculations show harmonic structural stability at 0 K, they do not rule out the possibility of structural changes induced by thermal fluctuations at finite temperatures. 
Thus, to show robust thermal stability of type-II GeSe$_2$ structures at room temperature (300 K), we performed the \textit{ab initio} molecular dynamics (AIMD) simulation~\cite{PhysRevLett.55.2471}. 
Using the ($3 \times 3 \times 3$) supercell to include $27$ formula units, AIMD was performed in the canonical ensemble (constant NVT). 
We controlled the temperature of the system by using the Nos\'{e}-Hoover thermostat~\cite{10.1063/1.447334, PhysRevA.31.1695}.
%Therefore, we further examined the thermodynamic stability of type-I and type-II GeSe$_2$ structures through canonical molecular dynamics (MD) simulations at room temperature (300 K). 
%
Figures~\ref{FIG5}(c) and (d) show the evolution of total potential energy and temperature over a 4-ps MD trajectory and includes snapshots of the final structures. 
Although small fluctuations in total potential energy are observed during the MD runs, no structural collapse or significant phase transformations occur, indicating that type-II GeSe$_2$ remains thermally stable over the simulated timescale at room temperature. 
Moreover, analysis of ensemble-averaged MD data confirms the achievement of thermal equilibrium within the simulation timeframe. 
Overall, these dynamic and thermodynamic stability analyses reinforce the viability of type-II GeSe$_2$ as a stable and efficient absorber material suitable for practical photovoltaic applications. 
These findings provide a robust theoretical basis to guide future experimental validation and optimization of GeSe$_2$-based solar cells.

\section{Conclusion}
%\addsj{SJK:논문의 본문에서 type-I은 unstable한 것을 보였으므로, 결론에서는 type-II로 한정하여 결론을 내리는게 좋지 않을까요?}

%
In conclusion, our computational study identifies bulk-stacked type-II GeSe$_2$ as a promising photovoltaic absorber material, highlighting its superior optoelectronic properties and enhanced absorption. Utilizing $GW$+BSE calculations, we quantified the quasiparticle band gaps and excitonic effects, demonstrating that type-II GeSe$_2$ achieves an SLME of approximately 25.6\% at 0.5 $\mu$m absorber thickness. This efficiency is comparable to or superior to that of the leading emerging 1D photovoltaic materials stacked in bulk, such as Sb$_2$Se$_3$ and Sb$_2$S$_3$. Crucially, phonon dispersion calculations and room-temperature molecular dynamics simulations confirm that type-II GeSe$_2$ phases are dynamically stable and thermodynamically robust, suitable for practical solar-cell applications. These findings underscore the promising potential of GeSe$_2$-based nanostructures as next-generation photovoltaic materials and establish a theoretical foundation for subsequent experimental validation and device integration.

%
%\addsj{SJK:정도를 나타내는 수식어를 잘 안쓰는 편이라서 우선은 줄을 그어놓았습니다만 필요하시면 다시 살리셔도 좋습니다.}

%
%\addsj{SJK:type-I가 결과적으로 unstable한데, photovoltaic 결과를 보여주는게 필요할지요? stability를 앞으로 옮길까요?}
\clearpage

\section*{Acknowledgement}
%
%S.-H.K was supported by Korea Institute of Science and Technology Information (KISTI) (K26L4M2C2). 
S.~Kim was supported by the National Research Foundation (NRF) of Korea (Grant no.~NRF-2022R1F1A1074670, ~RS-2024-00410027 and RS-2025-16065427).
B.G.J was supported by the National Research Council of Science \& Technology(NST) grant by the Korea government (MSIT) (No. CAP25061-000) and the National Supercomputing Center with supercomputing resources, including technical support (KSC-2025-CRE-0367).
Y.K. was supported by the National Research Foundation of Korea (NRF) grant funded by the Korea government (MSIT) (Grant no. RS-2025-00553820).
Authors thank the computational support from the Center for Advanced Computation (CAC) at Korea Institute for Advanced Study (KIAS).

\section*{Conflicts of interest}
There are no conflicts to declare.

\bibliographystyle{apsrev4-2}% your bst file here
\bibliography{biblio}

\end{document}